\documentclass[a4paper,10pt]{revtex4-2}
\usepackage{mathrsfs}
\usepackage{graphicx}
\usepackage{latexsym}
\usepackage{amsmath}
\usepackage{amssymb}
\usepackage{textcomp}
\usepackage{amsbsy}
\usepackage{graphics}
\usepackage{epstopdf}
\usepackage{color}

\allowdisplaybreaks[4]

\begin{document}

\tolerance=5000

\title{Holographic description  of the dissipative unified dark fluid model with axion field   }

\author{I.~Brevik,$^{1}$\,\thanks{iver.h.brevik@ntnu.no}
A.~V.~Timoshkin,$^{2,3}$\,\thanks{alex.timosh@rambler.ru}
}
 \affiliation{ $^{1)}$ Department of Energy and Process Engineering,
Norwegian University of Science and Technology, N-7491 Trondheim, Norway\\
$^{2)}$Institute of Scientific Research and Development, Tomsk State Pedagogical University (TSPU),  634061 Tomsk, Russia \\
$^{3)}$ Lab. for Theor. Cosmology, International Centre  of Gravity and Cosmos,  Tomsk State University of Control Systems and Radio Electronics
(TUSUR),   634050 Tomsk, Russia
}

\tolerance=5000

\begin{abstract}
In this article we extend an axion F(R) gravity model, and apply the holographic principle to describe in a unifying manner the early and the late-time universe when the general equation of state (EoS) contains a bulk viscosity. We assume a  spatially flat Friedmann-Robertson-Walker (FRW) universe model. We use a description based on the generalized infrared-cutoff holographic dark energy  proposed by Nojiri and Odintsov (2006, 2017), and explore the evolution of the universe when the EoS describes the asymptotic behavior between the dust in the early universe and the late universe. We explore various forms of the bulk viscosity, and   calculate analytical expressions for the infrared cutoffs in terms of the particle horizon. In this way we  obtain a unifying description of the early and the late-time universe in the presence of axion matter,  via a viscous holographic fluid model.


\end{abstract}

\date{\today}
\maketitle

\section{Introduction}

The holographic principle [1] is one of the current  approaches aiming  for   describing  the evolution of the universe. A generalized form of the  cutoff holographic dark energy (HDE) model was proposed Nojiri and Odintsov [2, 3]. The model  can be  applied to give a unified description of the early and the late-time accelerated universe. The theory of the holographic principle is associated with the thermodynamics of black holes and string theory [4, 5]. The infrared cutoff can be represented as a combination of various FRW universe parameters: the Hubble function, the particle and the future event horizons, the cosmological constant, and the finite life time of the universe. In the general case, infrared cutoff may be constructed as an arbitrary combination all above quantities and their derivatives. If the life time of the universe is finite due to singularities of various types,  the infrared radius depends also on the singularity time. Various versions of the holographic cutoffs were considered in [6–13]. Earlier,  as was shown in [14-16],  all the known holographic dark energy models represent subclasses of the Nojiri-Odintsov HDE. The holographic theory of the universe is well confirmed by astronomical observations [17–22].
In the present article we describe both the early-time and the late-time cosmic accelerating expansion  in a single cosmological model. In Ref.~[23],  a unifying approach was proposed to describe both the early-time and the late-time universe based on phantom cosmology. A unified model of  dark energy and dark matter  in  standard FRW cosmology was suggested in the works [24-27]. We will here suppose that the main component of cold dark matter in the universe is  axion matter [28-34].

\section{	Holographic Description of the accelerated universe with axion matter}

Let us highlight  the main aspects  of the holographic principle, following the terminology given by Li in [1]. The holographic principle states that all physical quantities within the universe, including the dark energy density, can be described by fixing some quantities at the boundary of the universe [29]. In the holographic description the main component in this context is the cutoff radius of the horizon. According to the generalized model referred to earlier  [2], the holographic energy density is taken to be inversely proportional to the squared infrared cutoff
 \begin{equation}
 \rho_{\rm hol}= 3c^2k^2L_{\rm IR}^{-2}
 \end{equation}						
where $k$  is Einstein’s gravitational constant, $G$ is  Newton’s gravitational constant, and $c>0 $ is a nondimensional constant. If the dark energy is described in this manner, it suggests that the horizon cutoff radius corresponds  to the infrared cutoff.
Although there is no concrete recipe  about how to choose
 this   parameter the  most suitable choice of it,  describing the accelerated expansion of the universe, is to equal it to  the particle horizon,  eventually to the   future horizon, defined respectively by [3]
 \begin{equation}
 L_p(t)= a(t)\int_0^t\frac{dt'}{a(t')},  \quad L_f(t)= a(t)\int_t^\infty \frac{dt'}{a(t')},
 \end{equation}
where $a(t)$ is the scale factor.
It should be noted  that not all choices of the  cutoff infrared radius  lead to an accelerated expansion of the universe. The choice of a cutoff radius is not arbitrary.  One way of obtaining a useful description of the era of inflation and the era of dark energy,  is to introduce a model of axion F(R) gravity. Within this model, it is possible to combine  the era of inflation with the era of dark energy.  We will consider a F(R) gravity model in the presence of a misalignment axion canonical scalar field $\phi$ with the approximate scalar potential
$V(\phi) \approx \frac{1}{2}m_a^2\phi_i^2,$
where $m_a$    is the axion mass and  $\phi_i$  is the axion scalar.
First of all, we will show how it is possible to consider the axion scalar  as constituting a cold dark matter perfect fluid.

Let us  consider the canonical equation of motion for the axion with scalar potential [30]
\begin{equation}
\ddot{\phi}+3H\dot{\phi}+m_a^2\phi=0.
\end{equation}

Since the second term in this equation describes  friction, we have to do with  decaying oscillations. The evolution of the universe describes a damped oscillator that approximately begins when  $H \sim m_a$  and lasts until $ H \gg m_a$.
Let us suppose that the oscillatory solution of the axion equation (3) has the form
\begin{equation}
\phi(t)= \phi_iA(t)\cos m_at,
\end{equation}
where  $\phi_i$  is the initial value of the axion field after the end of inflation and $A(t)$  is a slow-varying function. The function   is monotonous due to the conditions
\begin{equation}
\frac{A}{m_a} \sim \frac{H}{m_a} \approx \varepsilon \ll 1,
\end{equation}                                                     (5)
which are valid at cosmic times for which  $H \gg m_a$.
Using (4) and to conditions (5) one obtains  the  equation of motion (3) in the form
\begin{equation}
\frac{dA}{A}= -\frac{da}{a},
\end{equation}
which has the solution
\begin{equation}
A \sim a^{-3/2}.
\end{equation}
Let us  write the expressions for the energy density and pressure of  the axion field. They are equal to
\begin{equation}
\rho_a = \frac{1}{2}\dot{\phi}^2+ V(\phi)
\end{equation}
\begin{equation}
P_a =\frac{1}{2}\dot{\phi}^2-V(\phi),
\end{equation}
showing that  the misaligned  axion field can be considered as a canonical scalar field.

Calculating the term $\frac{1}{2}\dot{\phi}^2$   with the approximation (5), we obtain
\begin{equation}
\frac{1}{2}\dot{\phi}^2 \approx  \frac{1}{2}m_a^2\phi_i^2  A^2 \sin^2 m_at.
\end{equation}
Then the axion potential is equal to
\begin{equation}
V(\phi)= \frac{1}{2}m_a^2\phi_i^2  A^2 \cos^2 m_at.
\end{equation}
Using equations (10), (11) and the formula for the axion energy density (8), we obtain
\begin{equation}
\rho_a \approx \frac{1}{2}(m_a\phi_i A)^2.
\end{equation}
In this case, taking into account (7), the expression for the axion energy density becomes [29]
\begin{equation}
\rho_a \approx \frac{1}{2}\rho_m^{(0)}a^{-3},
\end{equation}
where $\rho_m^{(0)}= \frac{1}{2}(m_a\phi_i)^2$.

In summary, we  conclude that the axion energy density behaves as $\rho_a \sim a^{-3}$   for all cosmic times, assuming that $m_a \gg H$. Hence,   we have seen that the   axion scalar  can be considered as the constituent of a cold dark matter perfect fluid.

Let us  also calculate the pressure of the axion scalar, using (10, 11) with the help of (9),
\begin{equation}
P_a \approx -(m_a\phi_iA)^2\cos 2m_at.
\end{equation}
Writing the equation-of-state thermodynamic parameter for the axion scalar as $\omega_a = P_a/\rho_a$, we obtain
\begin{equation}
\omega_a = -\cos 2m_at,
\end{equation}
whose average value is zero. This is a consequence of our model of the axion being a cold dark matter particle.

\section{	Dissipative unified dark fluid model}

Let us consider the universe filled with viscous dark fluid in presence of  axion matter in a homogeneous and isotropic spatially flat Friedmann-Robertson-Walker (FRW) metric,
\begin{equation}
ds^2= -dt^2+\sum_{i=1}^3 (dx^i)^2.
\end{equation}
The modified Friedmann equation is [29]
\begin{equation}
H^2= \frac{1}{3}k^2(\rho+\rho_a)
\end{equation}
where  $H = \dot{a}/a$  is the Hubble function and  $\rho$   is the holographic dark energy.

We will describe the system, contains a viscous dark fluid in presence of the axion matter in terms of the parameters appearing in the effective inhomogeneous (EoS) in flat FRW space-time [35, 36]
\begin{equation}
p= \omega(\rho,t)\rho+f(\rho)- 3H\zeta (H,t).
\end{equation}
where  $\omega (\rho,t)$ is the thermodynamic parameter and $\zeta (H,t)$  is the bulk viscosity, which in general depends on both the Hubble function and on the time t. From thermodynamic considerations we take the bulk viscosity to be positive .

Let us consider the model of a unified description of the early and late universe. For this purpose, we choose the function in the form [25]
\begin{equation}
f(\rho)= \frac{\gamma \rho^n}{1+\delta \rho^m}.
\end{equation}
where  $\gamma, \delta, n, m$  are free parameters.
The function $f(\rho)$   in the formula (19) of the EoS provides the description of the unified early and late universe.  Using interpolation between different powers in the expression (19), we can be describe the asymptotic behavior between the dust in the early universe and the late universe [26].

Dissipative processes are  described with the bulk viscosity in the form [36]
\begin{equation}
\zeta(H,t)= \xi_1(t)(3H)^p,
\end{equation}
where the parameter $p$ is positive.

The energy conservation law takes the standard form
\begin{equation}
\dot{\rho}+3H(\rho+p)=0.
\end{equation}
We will now distinguish between two cases.

\noindent {\bf Case 1.}

\bigskip

First,  we consider the simplest case, when the thermodynamic parameter  is $\omega =\omega_0$   and  the bulk viscosity  is $\zeta(H,t)=\zeta_0$,  both constants. We restrict ourselves to the values  $n=\frac{3}{2}$ and $m=1$, what corresponds to $n-m=\frac{1}{2}$.
Then the equation of state (18) will read
\begin{equation}
p= \omega_0\rho + \frac{\gamma \rho^{3/2}}{1+\delta \rho^{1/2}} -3\zeta_0H.
\end{equation}
The Friedmann equation (17), when the axion energy density (13) is taken into account, reads
\begin{equation}
\rho= \frac{3}{k^2}H-\frac{\rho_m^{(0)}}{a^3}.
\end{equation}
Using (22), (23) in the approximation of large $\rho$ one obtains from (21)
\begin{equation}
\frac{2}{k^2}\dot{H}+\frac{3}{k^2}\left( \omega_0+\frac{\gamma}{\delta}+1\right)H^2-\left(\omega_0+\frac{\gamma}{\delta}\right)
\frac{\rho_m^{(0)}}{a^3}-3\zeta_0H=0.
\end{equation}
We write this equation in terms of the scale factor as
\begin{equation}
\ddot{a}a+\left[ \frac{3}{2}\left( \omega_0+\frac{\gamma}{\delta}\right) -\frac{1}{2}\right]\dot{a}^2-\frac{3}{2}\zeta_0k^2\dot{a}a-
\frac{1}{2}k^2\rho_m^{(0)}\left(\omega_0+\frac{\gamma}{\delta}\right)\frac{1}{a}=0.
\end{equation}
If we take $\omega_0=-\frac{1}{3}-\frac{\gamma}{\delta}$, equation (25) simplifies to
\begin{equation}
\ddot{a}-\frac{3}{2}\zeta_0k^2\dot{a}+\frac{k^2}{6}\rho_m^{(0)}\frac{1}{a^2}=0.
\end{equation}
When the viscosity $\zeta_0 \rightarrow 0$ the solution of (26) becomes
\begin{equation}
\frac{1}{C_1}\sqrt{ a\left( C_1a+\frac{1}{3}k^2\rho_m^{(0)}\right)} +
\frac{1}{3C_1^{3/2}}k^2\rho_m^{(0)}
\ln \Big|\frac{\sqrt{3C_1a}}{k\sqrt{\rho_m^{(0)}}} \left( 1
- \sqrt{\frac{1}{3C_1}k^2\rho_m^{(0)}\frac{1}{a}+1}\,\right)\Big|=t+C_2,
\end{equation}
where $C_t \neq 0$ and $C_2$ is arbitrary.

Let us consider the particular case when $C_1=C_2=0$. The the solution of the equation of motion becomes
\begin{equation}
a(t)=\rho_a^{(0)} t^{\frac{2}{3}},
\end{equation}
where $\rho_a^{(0)}= \left( \frac{1}{3}k^2\rho_m^{(0)}\right)^{1/3}$.

Correspondingly, the Hubble function becomes
\begin{equation}
H(t)= \frac{2}{3t},
\end{equation}
and the particle horizon $L_p$ becomes
\begin{equation}
L_p= 3t.
\end{equation}

Let us  now interpret equation (26) from a holographic point of view.
From [2],  the Hubble function $H$    can be expressed in terms of the particle horizon and its time derivative,
\begin{equation}
H=\frac{\dot{L}_p-1}{L_p},
\quad \dot{H}= \frac{\ddot{L}_p}{L_p}-
\frac{\dot{L_p}^2}{L_p^2}+
 \frac{\dot{L}_p}{L_p^2}.
\end{equation}
In our case it is necessary to express the scale factor and its time derivative in terms of the particle horizon and its derivative,
\begin{equation}
a= \rho_a^{(0)}\left( \frac{\dot{L}_p}{L_p}\right)^{- \frac{2}{3}}, \quad \dot{a}=\frac{2}{3}\rho_a^{(0)}\left( \frac{\dot{L}_p}{L_p}\right)^{\frac{1}{3}}, \quad \ddot{a} = -\frac{2}{9}\rho_a^{(0)}\left( \frac{\dot{L}_p}{L_p}\right)^{\frac{4}{3}}.
\end{equation}
Thus, by using (32) we can rewrite the energy conservation equation (26) in the holographic form
\begin{equation}
\rho_a^{(0)}\left(\frac{2}{9}\frac{\dot{L}_p}{L_p}+\zeta_0k^2\right) - \frac{k^2}{6}\frac{\rho_m^{(0)}}{\rho_a^{(0)}}\left(\frac{\dot{L_p}}{L_p}\right)^{\frac{1}{3}}=0.
\end{equation}
Thereby, we have applied the holographic principle to this model. Equation (33) shows the holographic description of the viscous dark fluid model with axion matter.

{ \bf Case 2.}

\bigskip

Next, we will consider the case, when the thermodynamic parameter  is constant,   and assume the bulk viscosity to be linearly proportional to the Hubble function,   where the viscosity parameter   is positive (in natural units where the fundamental length is cm, the dimension of   is , and since the dimension of    is , the dimension of   is ).
We will work in the approximation for the parameters  and , considered in the previous case.
The EoS (18) becomes
\begin{equation}
p= \omega_0\rho +
\frac{\gamma \rho^{3/2}}{ 1+\rho \delta^{1/2} } -9\tau H^2.
\end{equation}
Using (21), (23), (34) in the approximation for large   one obtains the differential equation of motion
\begin{equation}
\frac{2}{k^2}\dot{H}
 +\frac{3}{k^2}\left(\omega_0+\frac{\gamma}{\delta}-9\tau +1\right)H^2
- \left(\omega_0+\frac{\gamma}{\delta}\right)\frac{\rho_m^{(0)}}{a^3}=0.
\end{equation}
Let us  rewrite this equation in terms of the scale factor,
\begin{equation}
\ddot{a}a +\frac{1}{2}\left[ 3\left(\omega_0+\frac{\omega}{\delta}\right)  -9\tau k^2+1 \right]\dot{a}^2 -\frac{1}{2}k^2\rho_m^{(0)}\left(\omega_0+\frac{\gamma }{\delta}\right)\frac{1}{a} =0.
\end{equation}
Let us take $\omega_0 = -\gamma/delta$,  and then obtain the description through viscous dark fluid in the absence of axion dark matter.
The equation simplified and takes the form
\begin{equation}
\ddot{a}a + \frac{1}{2}(1-9\tau k^2)\dot{a}^2=0.
\end{equation}
The solution of (37) becomes
\begin{equation}
\frac{1}{b+1}a^{b+1}=(C_1t+C_2), \quad b \neq -1,
\end{equation}
where $b= \frac{1}{2}(1-9\tau k^2)$,  					(38)
where  $C_1$  and $C_2$  arbitrary constants.
If , the value of the viscous parameter , we obtain the solution of (37) in the form
\begin{equation}
a(t)= C_2e^{C_1t}.
\end{equation}						
Next, we calculate the particle horizon
\begin{equation}
L_p= \frac{1}{C_1}\left(e^{C_1t}-1\right).
\end{equation}
Using, (39) and (40) we express the scale factor in terms of the particle horizon, its derivatives
\begin{equation}
a= C_1(C_1L_p+1), \quad \dot{a}= C_1C_2\dot{L}_p, \quad \ddot{a}= C_1C_2 \ddot{L}_p,
\end{equation}
then the holographic representation of the motion equation (37) is
\begin{equation}
(1+C_1L_p)\ddot{L}_p+\frac{1}{2}(1-9\tau k^2)\dot{L}_p^2=0.
\end{equation}
The equation (42) represents a reconstruction of the energy conservation equation, according to the holographic principle. Thus, we applied the holographic principle in the unified dissipative dark fluid model to obtain the appropriate energy conservation law.

	\section{ Conclusion}

In the present paper we have considered a unified model of the early and the late-time universe, in a homogeneous and isotropic spatially flat Friedmann-Robertson-Walker metric, from a holographic point of view. We assumed that the universe is filled with viscous dark fluid in presence axion matter, and presented the energy conservation equation in a holographic language. We  showed the equivalence between viscous fluid cosmology and holographic fluid cosmology assuming the  cutoff model introduced by Nojiri and Odintsov [2, 3].

For this, we  identified the infrared radius $L_{\rm IR}$   with the particle horizon $L_p$. We assumed  a general EoS for the viscous dark fluid in the presence of axion matter. We  applied the holographic principle to cosmological models with constant value of the thermodynamic parameter $\omega(\rho,t)$    and considered different  forms of the bulk viscosity $\zeta(H,t)$.  In every model the infrared radius, in the form of a particle horizon, was calculated in order to obtain the energy conservation equation. Despite the fact that in the inflationary scenario the contribution of bulk viscosity is usually insignificant, with an influence increases only with the development of the universe, we  described the holographic picture using a viscous fluid. Thus, equivalence is established between the description of the unified model of the early and late universe with the help of a viscous dark fluid, and its holographic description based on a  selection of the infrared radius.

One may ask if  there is an  agreement between the holographic theory and astronomical observations. A comparative analysis was given in [37] examining  the  holographic dark energy model on the brane. The analysis was carried out for  relationships between apparent magnitudes and redshifts for distant supernova Ia, Hubble parameters for different redshifts, and baryon acoustic oscillations.  For a wide range of the parameters, the observational data were found to be  in good agreement with theoretical predictions.

\bigskip

\noindent {\bf Acknowledgment}

\noindent  This work was supported by Russian Foundation for Basic Research; Project No. 20-52-05009.

\end{document}